\documentclass{aa}
\usepackage{graphicx}
\begin{document}
%
   \title{Determination of proper motions and membership of the open star cluster NGC~2548\thanks{Table~5 is only available in electronic form from CDS via
    anonymous ftp 130.79.128.5, and by e-mail request to Lola.Balaguer@am.ub.es}
    }

   \author{Z.Y. Wu\inst{1,2,3}, K.P. Tian\inst{1,2,3}, L.
   Ba\-la\-guer-\-N\'u\-\~nez\inst{5,1}, C. Jor\-di\inst{5,6}, 
   J.L. Zhao\inst{4,1,2,3}
   and J. Guibert\inst{7}
          }

   \offprints{L. Balaguer-N\'u\~nez, \email{Lola.Balaguer@am.ub.es}}

   \institute{Shanghai Astronomical Observatory, CAS Shanghai 200030,
                P.R. China
         \and
            Chinese National Observatories, CAS, P.R. China
         \and
            Joint Lab of Optical Astronomy, CAS, P.R. China
         \and
            CCAST (WORLD LABORATORY) P.O. Box 8730, Beijing 100080,
            P.R. China
         \and
            Departament d'Astronomia i Mateorologia, Universitat de
            Barcelona, Avda. Diagonal 647.  E-08028 Barcelona, Spain 
         \and
            Institut d'Estudis Espacials de Catalunya - IEEC, Edif. Nexus
            Gran Capit\`a 2-4, E-08034 Barcelona, Spain
        \and
            Centre d'Analyse des Images, Observatoire de Paris,
            B\^atiment Perrault, 77 Av. Denfert-Rochereau,
            F-75014 Paris, France
             }
\date{Received ; accepted}

\authorrunning{Z.Y. Wu et al}
\titlerunning{Proper motions and membership of the cluster NGC~2548}

   \abstract{
   Absolute proper motions, their corresponding errors and membership
   probabilities of 501 stars in the intermediate-age open cluster NGC~2548 region
   are determined from MAMA\thanks{MAMA (Machine Automatique \`a Mesurer pour
   l'Astronomie, http://dsmama.obspm.fr) is developed and operated by INSU/CNRS
   and Observatoire de Paris.}
   measurements of 10 photographic plates. The plates
   have the maximum epoch difference of 82 years and they were taken
   with the double astrograph at Z\^ o-S\`e station of Shanghai
   Observatory, which has an aperture of 40 cm
   and a plate scale of 30\arcsec~mm$^{-1}$. The average proper motion
   precision is 1.18 mas~yr$^{-1}$. These proper motions are used to
   determine the membership probabilities of stars in the region.
   The number of stars with membership probabilities higher than 0.7
   is 165.
      \keywords{
       Galaxy: open clusters and associations: individual: NGC~2548 --
                astrometry -- Galaxy: kinematics and dynamics
               }
   }

   \maketitle

%

\section{Introduction}

    The open cluster NGC~2548, also known as M~48, in Hydra
    ($\alpha_{2000}$=$8^{\mathrm h}\- 13^{\mathrm m}\- 48^{\mathrm s}\-$, 
    $\delta_{2000}=-5{\degr}48\arcmin$)
    seems to be an intermediate-age open cluster with an estimated distance
    of 630 pc (Pesch \cite{pesch}) or 530 pc (Clari\'a \cite{Claria}). 
    It has not been the object of
    any recent complete astrometric or photometric study
    (Ebbinghausen \cite{Ebb}; Li \cite{Li}), in spite of being an extended
    object with an apparent dia\-me\-ter of $30\arcmin$ (Trumpler \cite{Trum}) or even 
     $54\arcmin$ (Collinder \cite{Coll}) and brilliant enough to be
    in the Messier list (s.XVIII) with number 48 (Messier \cite{mess}).
    It was even considered inexistent for several years 
    owing to the fact that Messier mistook its real coordinates owing to a 
    change of sign in its relative position
    with a final difference in declination of $5^{\circ}$.

      There is no
    feasible estimation of its age but it seems to be an intermediate-age
    open cluster, around $\log t$ = 8.5 (Lyng\aa \ \cite{lynga}),
    with a slightly poorer CN abundance
    than the giants of the Hyades but significantly
    richer than the K giants of the solar neighborhood (Clari\'a
    \cite{Claria}; Twarog et al. \cite{Twa}).

    Proper motions of bright stars in this region were published by
    Ebbinghausen (\cite{Ebb}) with four pairs of plates of only a
    maximum epoch difference of 28 years. Li (\cite{Li}) 
    pu\-bli\-shed
    positions and proper motions in the field of NGC~2548 based on
    three plates taken with the 40 cm astrograph at Shanghai
    Z\v o-S\`e station with a maximum epoch difference of only 14 years.

    In this paper we determine, for the first time, precise
    absolute proper motions of 501 stars within a 1\fdg6 $\times$
    1\fdg6 area in the NGC~2548 region, from automatic MAMA measurements
    of 10 plates, five of which are newly taken. The estimated 
    membership probabilities lead
    us to a complete astrometric study of the cluster area.
    Section 2  des\-cri\-bes the plate material and its measurement as well as 
    the proper motion reduction procedure and results with comparisons with the 
    Hipparcos and Tycho-2 catalogues. Section 3 accounts for the membership 
    determination. Section 4 is devoted to the analysis of results.
    Finally, a summary is presented in Section 5.


\section{Plate measurement and proper motion reduction}

\subsection{Plate material and measurements}

   Ten plates of the NGC~2548 region were available.
   They were taken with the double astrograph at the Z\^ o-S\`e station
   of Shanghai Observatory. This telescope, built by Gaultier in Paris
   at the beginning of the last century, has an aperture of 40 cm, a focal
   length of 6.9 m and hence a plate scale of 30\arcsec~mm$^{-1}$. The size of the
   old plates is 24 cm by 30 cm, or 2\fdg0 $\times$ 2\fdg5,
   and that of the new ones is 20 cm by 20 cm, or 1\fdg65 $\times$ 1\fdg65.

   The oldest
   plate was taken in 1916, and the newest ones in 1998. Relevant
   information on these plates is shown in Table~\ref{plates}. The hour
   angles are not provided in Table~\ref{plates} because the
   starting exposure time of the old plates was not recorded.

\begin{table*}
\leavevmode
\caption {Plate material}
{\scriptsize
\begin {tabular} {ccccccc}
\hline
   Plate  & Epoch & Exp.Time & Plate center & Plate size & n. of  \\
  Id. & (1900+) & min  & ($\alpha_{\mathrm J2000}$ $\delta_{\mathrm J2000}$)& cm  & stars \\
\hline
     CL422    &  16.03.24  &  90 & 8\fh2283 $-5$\fdg779 & 24$\times$30 & 450 \\
     CL534    &  30.03.21  &  90 & 8\fh2286 $-5$\fdg775 & 24$\times$30 & 556 \\
     CL535    &  30.03.28  &  90 & 8\fh2291 $-5$\fdg767 & 24$\times$30 & 566 \\
     CL56006  &  56.03.14  &  60 & 8\fh2288 $-5$\fdg779 & 24$\times$30 & 577 \\
     CL56007  &  56.03.16  &  60 & 8\fh2372 $-5$\fdg779 & 24$\times$30 & 558 \\
     CL98004  &  98.04.03  &  30 & 8\fh2447 $-5$\fdg958 & 24$\times$30 & 548 \\
     CL98047  &  98.12.16  &  30 & 8\fh2227 $-5$\fdg796 & 20$\times$20 & 268 \\
     CL98Tian &  98.12.25  &  30 & 8\fh2255 $-5$\fdg804 & 20$\times$20 & 268 \\
     CL98Chen &  98.12.25  &  30 & 8\fh2255 $-5$\fdg804 & 20$\times$20 & 298 \\
     CL98Gu   &  98.12.16  &  30 & 8\fh2227 $-5$\fdg796 & 20$\times$20 & 432 \\

\hline
\label{plates}
\end {tabular}
}
\end {table*}

    All plates were measured at the Centre d'Analyse des Images at
    the Observatoire de Paris, using the high precision
    microdensitometer "Machine Automatique \`a Mesurer pour l'Astronomie"
    (MAMA).
    This device has a superb optical and mechanical performance
    (Guibert et al.  \cite{Gui}). It uses a quartz-iodine illuminating
    source, whose transmitted light is detected by a reticon, 1024
    pixels large, with a pixel size of 10 ${\mu}$m, and the absolute
    accuracy of the measurements is 0.6 ${\mu}$m (Soubiran \cite{Sou}).
    After the plates were scanned, the resulting images
    were stored in a grid of $19\times 19$ sub-images for each plate.
    Once every plate was digitized, we 
    identified all point sources in these 361 frames. The source
    extraction was performed on each frame using SExtractor
    (Bertin \& Arnouts \cite{Ber}), a software dedicated to the
    automatic analysis of astronomical images using a
    multi-threshold algorithm allowing good object deblending.
    To improve the accuracy of the measurements, we chose to
    retain only rather bright objects with a signal-to-noise ratio of at
    least 12.

    During the scanning, MAMA includes in the
    catalogue not only real images, but also the plate grid,
    emulsion flaws, plate annotations and scratches. A first step
    is required to reject spurious detections, most of which
    can be recognized with analysis of the object shape,
    by removing the plate grid, annotations and big scratches by
    visual comparison with the original plates. 
    The shape parameters given by SExtractor can be
    used to clarify the remaining stars but the most direct way to reject
    spurious detection is by comparison between plates,
    searching each object from one plate to the other, and
    retaining only the paired data.
    There is a total of 596 stars measured with a
    limiting magnitude $B_T$ around 14.
    This limiting magnitude was roughly estimated from the stars
    in common with Tycho-2 catalogue. 

    The detection of 182 PPM and ACT stars on the whole plate
    allowed us to determine the star positions in a common
    system. Thus for each scanned plate, we obtained a
    preliminary astrometric catalogue to perform
    initial cross-identification between plates.

\subsection{Proper motions}

    The absolute proper motions for 596 stars in
    the region of NGC~2548 were reduced on the basis of the MAMA
    measurements following the central overlapping procedure
    (Russel \cite{russel}; Wang et al.\ \cite{Wang95}, \cite{Wang96},
    \cite{Wang00}).
    This method determines the plate-to-plate
    transformation parameters, the star motions and their errors simultaneously. 
    Stellar positions and absolute proper motions
    were reduced from a catalogue used as the original data
    for the first iteration. As initial catalogue, 265 stars from
    the Tycho-2 ca\-ta\-lo\-gue at epoch J2000 (H\a{o}g et al.\ \cite{tyc2a})
    were chosen on the basis of the results of the PPM and ACT
    as\-tro\-me\-tric catalogue given by MAMA.
    To select the best plate constant model, we used Eichhorn \& Williams' criterion 
    (Eichhorn \& Williams \cite{Eicwill}, Wang et al.\ \cite{Wang82}) and obtained 
    a model with
    six linear constants on coordinates, a
    magnitude and a coma term, and a 
    magnitude distortion
    term. Magnitudes used were the instrumental magnitudes. 
    All the proper motions are constrained by having at least one measurement 
    from the modern epoch plates, i.e. taken in 1998. 
    The whole
    process is iterated until the re\-sul\-ting proper motions
    converge. 
    We iterate the
    process until
    mean differences in position are smaller than 1.1 mas, the r.m.s. smaller
    than 3.6 mas and the differences in proper motion below 0.1 mas~yr$^{-1}$,
    yielding a final outcome of 501 stars.

\begin{table*}
\leavevmode
\caption {Mean precisions of proper motions as a function of the     
          number of plates in the NGC~2548 region. 
          (Units are mas~yr$^{-1}$.) $N$ is the number of stars}
\begin {tabular} {ccccc}
\hline
\hline
 $plates$ & $N$ & $\epsilon_{\mu_{\alpha}\cos\delta}$ & $\epsilon_{\mu_{\delta}}$ & $\epsilon_{\mu}$ \\
\hline
3  &   4  &  5.29 &  2.90 &  6.30 \\
4  &   8  &  2.92 &  3.12 &  4.44 \\
5  &  46  &  1.76 &  1.11 &  2.19 \\
6  & 119  &  1.19 &  0.89 &  1.53 \\
7  &  27  &  1.15 &  0.76 &  1.42 \\
8  &  20  &  0.90 &  0.70 &  1.17 \\
9  &  50  &  0.74 &  0.55 &  0.94 \\
10 & 199  &  0.59 &  0.46 &  0.77 \\
\hline
\hline
\end {tabular}
\label{error}
\end {table*}

\begin{figure}
\begin{center}
\resizebox{6cm}{!}{\includegraphics{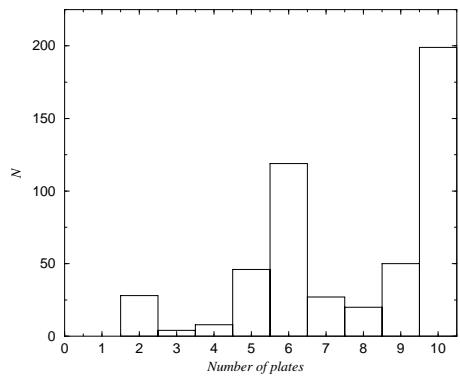}}
\end{center}
\caption{The number of stars $vs$ the number of available plates}
\label{nplpr}
\end{figure}

    Table~\ref{error} shows mean precisions of final proper motions
    for stars in the NGC~2548 region detected on different numbers of measured
    plates (greater than 2).
    The units of the proper motions and their precisions all along are
     mas~yr$^{-1}$. The precision of the final proper motions strongly depends 
    on the number of plates. 
    Figure~\ref{nplpr} gives the number of stars for which various
    numbers of plates are available. More than $90\%$ of proper
    motions were obtained from at least 5 plates.

    The r.m.s. errors on proper motions for more than $90\%$ of stars
    are ${\epsilon}_{\mu_{\alpha}\cos\delta}$= 0.92 mas~yr$^{-1}$, 
    ${\epsilon}_{\mu_{\delta}}$ = 0.68 mas~yr$^{-1}$
    and ${\epsilon_{\mu}}$= 1.18 mas~yr$^{-1}$, where
    \begin{math}\epsilon_{\mu}=\sqrt{{\epsilon ^{2}_{\mu_{\alpha}\cos\delta}+\epsilon^{2}_{\mu_{\delta}}}} 
    \end{math}. In the most precise case, the r.m.s. errors are 
    0.77 mas~yr$^{-1}$ for stars with 10 plates ($40\%$ of stars).

\begin{figure*}
\begin{center}
\resizebox{17cm}{!}{\includegraphics{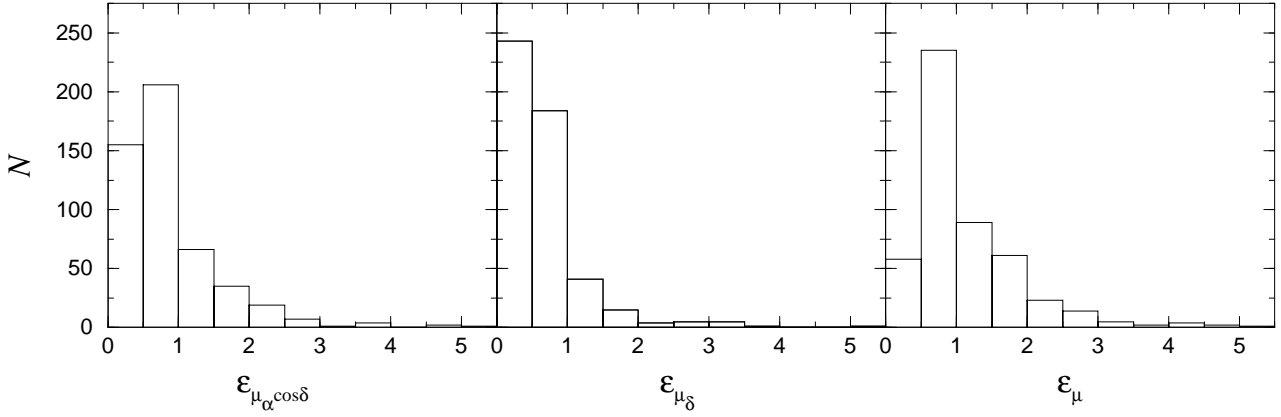}}
\end{center}
\caption{The number of stars $vs$ the r.m.s. errors on proper motions in mas~yr$^{-1}$ }
\label{nerrhis}
\end{figure*}

    Figure~\ref{nerrhis} shows the distribution of r.m.s.\ proper motion errors with
    the number of stars: $N$
    {\it versus\/} $\epsilon _{\mu_{\alpha}\cos\delta}$, $\epsilon _{\mu_{\delta}}$ and
    $\epsilon_{\mu} $. Thus the precisions
    of the proper motions of stars in the region of NGC~2548 obtained
    in this study are relatively high, thanks to the quality of the stellar 
    images taken with
    the 40 cm double astrograph and the excellent positioning behavior of the
    MAMA scanning machine.

\begin{figure}
\resizebox{\hsize}{!}{\includegraphics{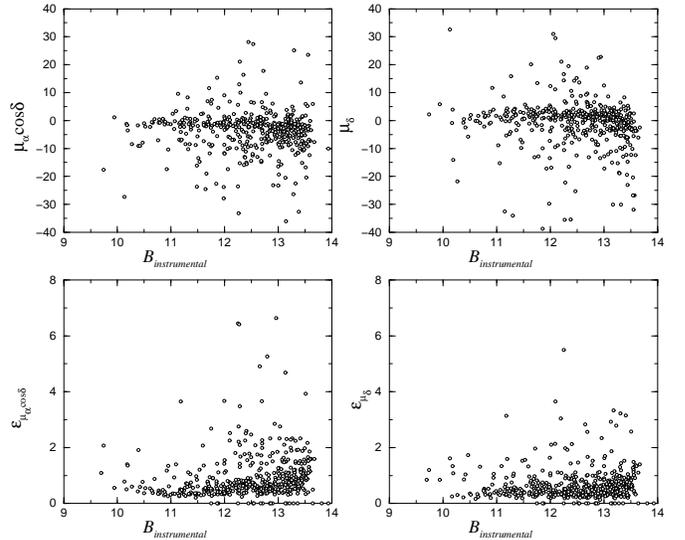}}
\caption{Proper motions (top) and their errors (bottom) $vs$ instrumental magnitude.
     Null errors are from proper motions calculated with only two plates. (Units are mas~yr$^{-1}$)}
\label{mag}
\end{figure}

    Figure~\ref{mag} gives $\mu_{\alpha}\cos\delta$, $\mu_{\delta}$ 
    and their errors with respect to
    instrumental magnitude. 

\begin{figure}
\begin{center}
\resizebox{\hsize}{!}{\includegraphics{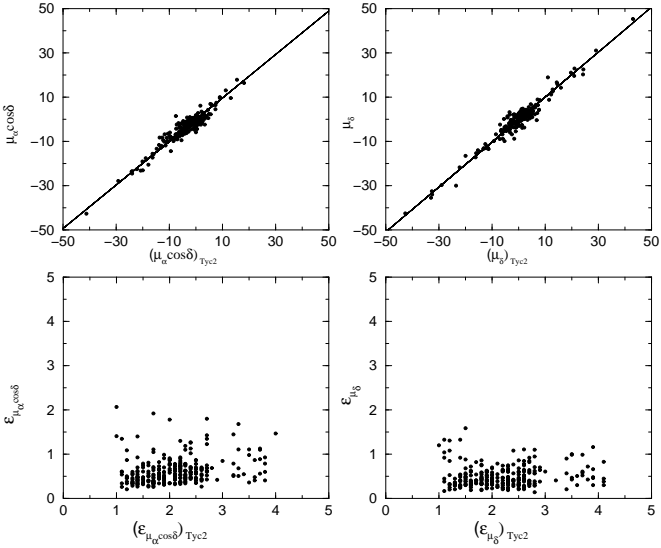}}
\end{center}
\caption{Proper motions and their errors from this paper compared to those 
in Tycho-2 catalogue. (Units are mas~yr$^{-1}$)}
\label{comp}
\end{figure}

    Our absolute proper motions and their errors are compared with those of 
    Tycho-2 catalogue
    in Fig.~\ref{comp}. We would like to highlight the
    precision of the proper motions derived in this paper. 
    Mean differences in the sense ours minus Tycho-2 are $-$0.123 ($\sigma$~=~2.112) 
    and $-$0.203 ($\sigma$~=~2.158) in $\mu_{\alpha}\cos\delta$ and $\mu_{\delta}$, respectively.
    No apparent systematic residuals were 
    found as a function of magnitude. 
    A linear fit to the proper motion data gives us:

    $\mu_{\alpha}\cos\delta$ = $0.187({\pm} 0.133) 
     +$~$0.982({\pm} 0.009)$~$\cdot$~$(\mu_{\alpha}\cos\delta)_ {\mathrm{Tyc2}}$;\ 
    $r$ = $0.988$ 

    $\mu_{\delta}$ = $0.201 \ ({\pm} 0.132)  + 1.016 \ ({\pm} 0.008)
     \cdot (\mu_{\delta})_{\mathrm{Tyc2}}$ ;\\
    $r$ = $0.992$ 

    \noindent  being $r$ the correlation coeficient.
	
    Only 8 stars were found in this region from the Hipparcos
    catalogue (ESA \cite{esa}). By comparing our absolute proper 
    motions with these common stars, the 
    mean differences are (in the sense ours minus Hipparcos) $-$0.610 ($\sigma$ = 1.710) and 
    $-$0.198 ($\sigma$ = 1.658) in $\mu_{\alpha}\cos\delta$ and $\mu_{\delta}$, respectively.
    We obtain the following linear fit:

    $\mu_{\alpha}\cos\delta$~=~$-0.458 ({\pm}0.756)+$~$1.005({\pm} 0.013)
     \cdot$~$(\mu_{\alpha}\cos\delta)_ {\mathrm{HIP}}$;\\
    $r$ = $0.999$ 

    $\mu_{\delta}$ = $-0.470 \ ({\pm} 0.694)  + 1.011 \ ({\pm} 0.014)
     \cdot (\mu_{\delta})_{\mathrm{HIP}}$ ;\\
    $r$ = $0.999$

\section{Membership determination}

    Accurate membership determination is essential for
    further astrophysical studies of clusters. The fundamental mathematical
    model set up by Vasilevskis et al.\ (\cite{vas}) and the technique
    based upon the maximum likelihood principle developed by Sanders
    (\cite{san}) have since been continuously refined.

    An improved method for membership determination of stellar
    clusters based on proper motions with different observed precisions was
    developed by Stetson (\cite{ste}) and Zhao \& He (\cite{zhaohe}).
    Zhao \& Shao (\cite{zhasha}) then added the correlation coefficient
    of the field star distribution to the set of parameters describing their
    distribution on the sky. This model has been frequently used 
    (Wang et al.\ \cite{Wang95}, \cite{Wang96}, \cite{Wang00}).

    We used a maximum likelihood method with a 9-parametric
    Gaussian model for the frequency function, as follows:

\begin{displaymath}
\Phi = \Phi_c + \Phi_f =
{n_c \cdot \phi_c
 + n_f \cdot \phi_f}, \nonumber
\end {displaymath}

\noindent where $\phi_c$, and $\phi_f$,
are the probability density functions of cluster members and
field stars respectively in the proper motion space,
with $n_c$ the normalized number of cluster stars, and $n_f$
the normalized number of field stars.

   The probability density function for the $i$-th star of the cluster can be written as
follows:

 \begin {eqnarray*}
 \phi_c(i) &=&\frac{1} {2\pi (\sigma_c^2+\epsilon_{(\mu_{\alpha}\cos\delta)_i}^2)^{1/2}
        (\sigma_c^2+\epsilon_{(\mu_{\delta})_i}^2)^{1/2}} \\
  & & \exp \left\{-\frac{1}{2}\left
        [\frac{\left[(\mu_{\alpha}\cos\delta)_{i}-(\mu_{\alpha}\cos\delta)_{c}\right]^2}
 {\sigma_c^2+\epsilon_{(\mu_{\alpha}\cos\delta)_i}^2}+\frac{[(\mu_{\delta})_{i}-(\mu_{\delta})_{c}]^2}
 {\sigma_c^2+\epsilon_{(\mu_{\delta})_{i}}^2}\right]\right\}, 
 \end {eqnarray*}

\noindent where [($\mu_{\alpha}\cos\delta)_{i}$ ,$(\mu_{\delta})_i$] are the 
proper motions of the $i$-th star, [$(\mu_{\alpha}\cos\delta)_{c},
(\mu_{\delta})_{c}$] the cluster mean proper motion,
$\sigma_c$ the intrinsic proper motion dispersions of member stars
and [$\epsilon_{(\mu_{\alpha}\cos\delta)_i}$,$\epsilon_{(\mu_{\delta})_{i}}$] the observed
errors on the proper motion components of the $i$-th star. \\

   Analogously, for the field we have:

\begin {eqnarray*}
\phi_f(i) =&\frac{1} {2\pi (1-\gamma^2)^{1/2}
      (\sigma_{(\mu_{\alpha}\cos\delta)_f}^2+\epsilon_{(\mu_{\alpha}\cos\delta)_i}^2)^{1/2}
      (\sigma_{(\mu_{\delta})_{f}}^2+\epsilon_{(\mu_{\delta})_{i}}^2)^{1/2}} \\
 &   \exp \left \{-\frac{1} {2(1-\gamma^2)}
 \left [\frac{[(\mu_{\alpha}\cos\delta)_{i} -(\mu_{\alpha}\cos\delta)_{f}]^2} 
    {\sigma_{(\mu_{\alpha}\cos\delta)_f}^2+\epsilon_{(\mu_{\alpha}\cos\delta)_i}^2}
                          \right. \right.  \\
                      &      \left. \left.
 -\frac {2\gamma[(\mu_{\alpha}\cos\delta)_{i}-(\mu_{\alpha}\cos\delta)_{f}]
[(\mu_{\delta})_{i}-(\mu_{\delta})_{f}]}
  {[\sigma_{(\mu_{\alpha}\cos\delta)_f}^2+\epsilon_{(\mu_{\alpha}\cos\delta)_i}^2]^{1/2}
[\sigma_{(\mu_{\delta})_{f}}^2+\epsilon_{(\mu_{\delta})_{i}}^2]^{1/2}}
 +\frac {[(\mu_{\delta})_{i}-(\mu_{\delta})_{f}]^2} 
{\sigma_{(\mu_{\delta})_{f}}^2+\epsilon_{(\mu_{\delta})_{i}}^2}\right] \right \} ,
\end {eqnarray*}

\noindent where 
[$(\mu_{\alpha}\cos\delta)_{f},(\mu_{\delta})_{f}$] are the field mean proper motion,
[$\sigma_{(\mu_{\alpha}\cos\delta)_f},\sigma_{(\mu_{\delta})_{f}}$] the field intrinsic 
proper motion dispersions and $\gamma$ is the correlation coefficient.\\


\section{Results and discussion}

   The unknown parameters for the assumed $\Phi$ distribution are [$n_c$,
$(\mu_{\alpha}\cos\delta)_{c}$, $(\mu_{\delta})_{c}$, $\sigma_c$] for the cluster and 
[$(\mu_{\alpha}\cos\delta)_{f}$, $(\mu_{\delta})_{f}$, $\sigma_{(\mu_{\alpha}\cos\delta)_f}$, 
$\sigma_{(\mu_{\delta})_f}$, $\gamma$] for the field po\-pu\-la\-tion. Membership probability
of the $i$-th star belonging to the cluster can be calculated
from

\begin {displaymath}
P_c(i)=\frac{\Phi_c(i)} {\Phi(i)}. {\hspace{3cm}}
\end {displaymath}

   By applying the standard maximum likelihood method, we obtained
the 9 distribution parameters and their corresponding uncertainties,
shown in Table~\ref{para}, where the units of the proper motions and proper
motion dispersions are mas~yr$^{-1}$.
    The quality of the fit should be optimized near the 
proper motion region occupied by the cluster stars, where the model is most crucial
to pro\-vi\-ding reliable membership determination. Outliers cause the estimated
distribution of field stars to be flatter than the actual one, thus increasing
the final probability of membership (Kozhurina-Platais et al.\  \cite{kozhu},
Cabrera-Ca\~no \& Alfaro \cite{canno}, Zhao \& Tian \cite{Zhao}, 
Zhao et al.\ \cite{Zhao82}).
To minimize the effect of high proper-motion field stars in the model, 
we restricted the membership determination to the range $|\mu| <$  30 mas~yr$^{-1}$.

\begin{table*}
\caption {Distribution parameters and their uncertainties
         for NGC~2548.
         The units of $\mu$ and $\sigma$ are mas~yr$^{-1}$ }
\begin {tabular} {lc c c c c c c c c c }
\hline

     & $n_c$
    & $ \mu_{\alpha}\cos\delta$ &
    $\mu_{\delta}$ & $\sigma_c$  & $\sigma_{\mu_{\alpha}\cos\delta}$ & $\sigma_{\mu_{\delta}}$ & $\gamma$\\
      & 
    &  &  &  &  &  &  &    \\
\hline
NGC~2548 & 0.382 &$-1.41$& 1.64 & 1.23 &  &  &     \\
    & $\pm0.025$ & $\pm0.12$ & $\pm0.12$ & $\pm0.08$ &  &  &    \\

field  &    
      & $-4.89$ & $-1.63$ &  & 7.37 & 8.21 & $-0.28$\\
   &   & $\pm0.07$ & $\pm0.57$ & &$\pm0.04$ &  $\pm0.29$ & 0.03 \\
\hline
\end {tabular}
\label{para}
\end {table*}

Table~5 lists the results for all 501 stars in the region of the
open cluster: column 1 is the ordinal star number; columns 2 and 3
are $\alpha_{\mathrm J2000}$ and $\delta_ {\mathrm J2000}$;
columns 4 and 6 are the respective absolute proper motions ($\mu_{\alpha}\cos\delta,
\mu_{\delta}$); columns 5 and 7 are the
standard errors of the proper motions; column 8 is the membership
probability of stars belonging to NGC~2548 $(P_c)$; column 9 is the
instrumental magnitude given by SExtractor;
column 10 gives the number of plates used and column 11 the identification number 
in Tycho-2 for the stars in common.

\begin{table*}
\caption {The cross-identification of stars in common 
         with the Hipparcos catalogue and the membership determination
         by Baumgardt et al.\ (\cite{Baum}). Comparison with the results 
         from this paper ($P_c$) gives complete agreement}
\begin {tabular} {cccccc}
\hline
Table~5 &  Hipparcos & Tycho-2 & BDA & Member & $P_c$ \\
\hline
 257&  40110 & 4859\_00078\_1 &   366 & N & 0.00 \\
 140&  40238 & 4859\_00250\_1 &  1005 & M & 0.87 \\
 139&  40254 & 4859\_00036\_1 &  1073 & M & 0.77 \\
 336&  40281 & 4859\_00921\_1 &       &   & 0.00 \\
 133&  40302 & 4855\_01706\_1 &  1320 & M & 0.78 \\
 234&  40348 & 4859\_01156\_1 &  1560 & M & 0.88 \\
 162&  40362 & 4859\_00092\_1 &  1628 & M & 0.84 \\
  42&  40498 & 4856\_00072\_1 &  2184 & N & 0.00 \\
\hline
\label{cross}
\end{tabular}
\end{table*}

    The cross-identification of our 8 common stars
    with the Hipparcos catalogue
    is shown in Table~\ref{cross}. Comparison with
    the membership determination calculated by Baumgardt et al.\ (\cite{Baum})
    for the seven stars in common shows complete agreement.

\begin{figure*}
\resizebox{17cm}{!}{\includegraphics{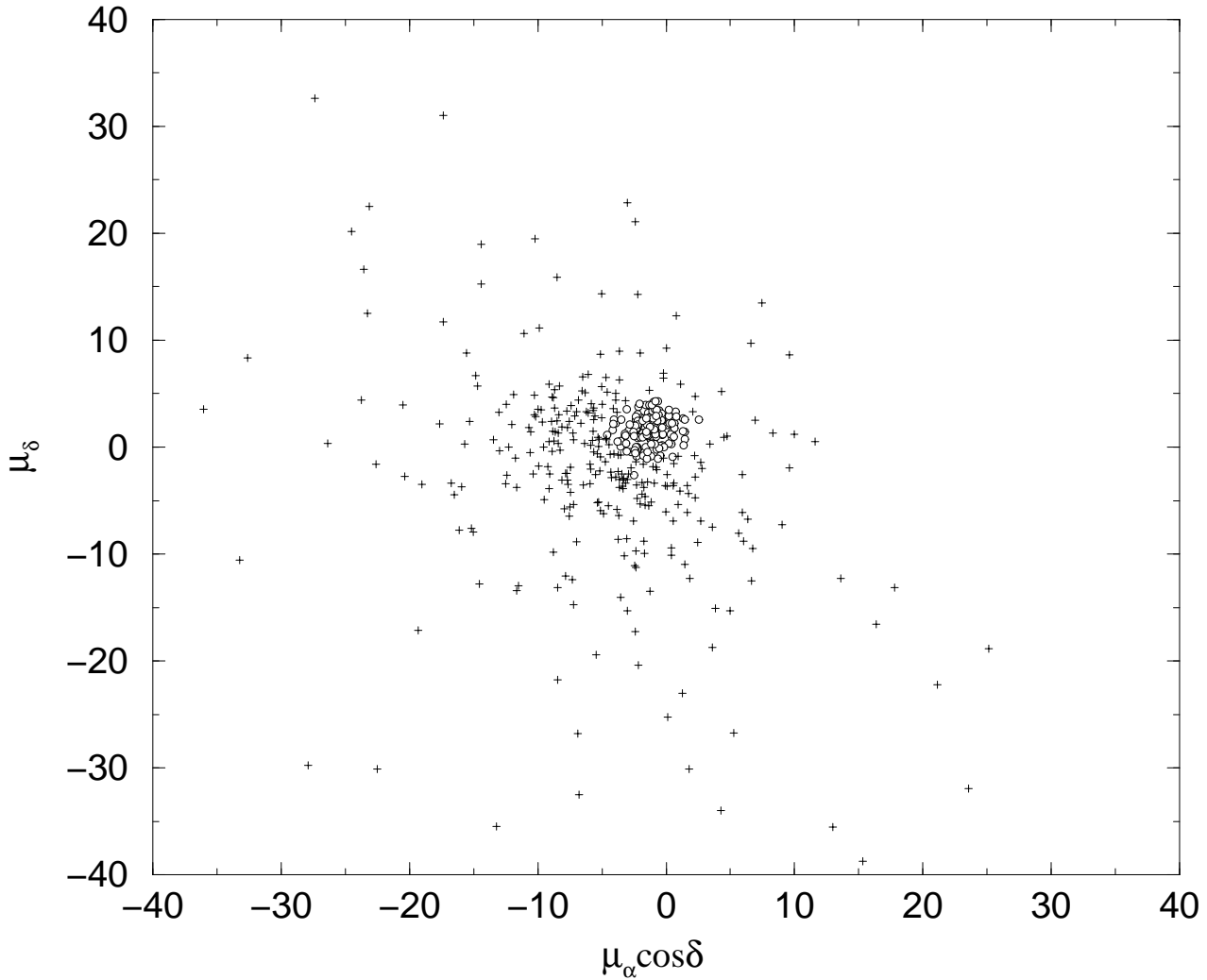}}
\caption{The proper motion vector-point diagram of NGC~2548. (Units are mas~yr$^{-1}$.)
           (``$\circ$" denotes a member of NGC~2548 with $P_c\ge0.7$,
            ``$+$" a field star with $P_c<0.7$)}
\label{mvpr}
\end{figure*}

\begin{figure*}
\resizebox{17cm}{!}{\includegraphics{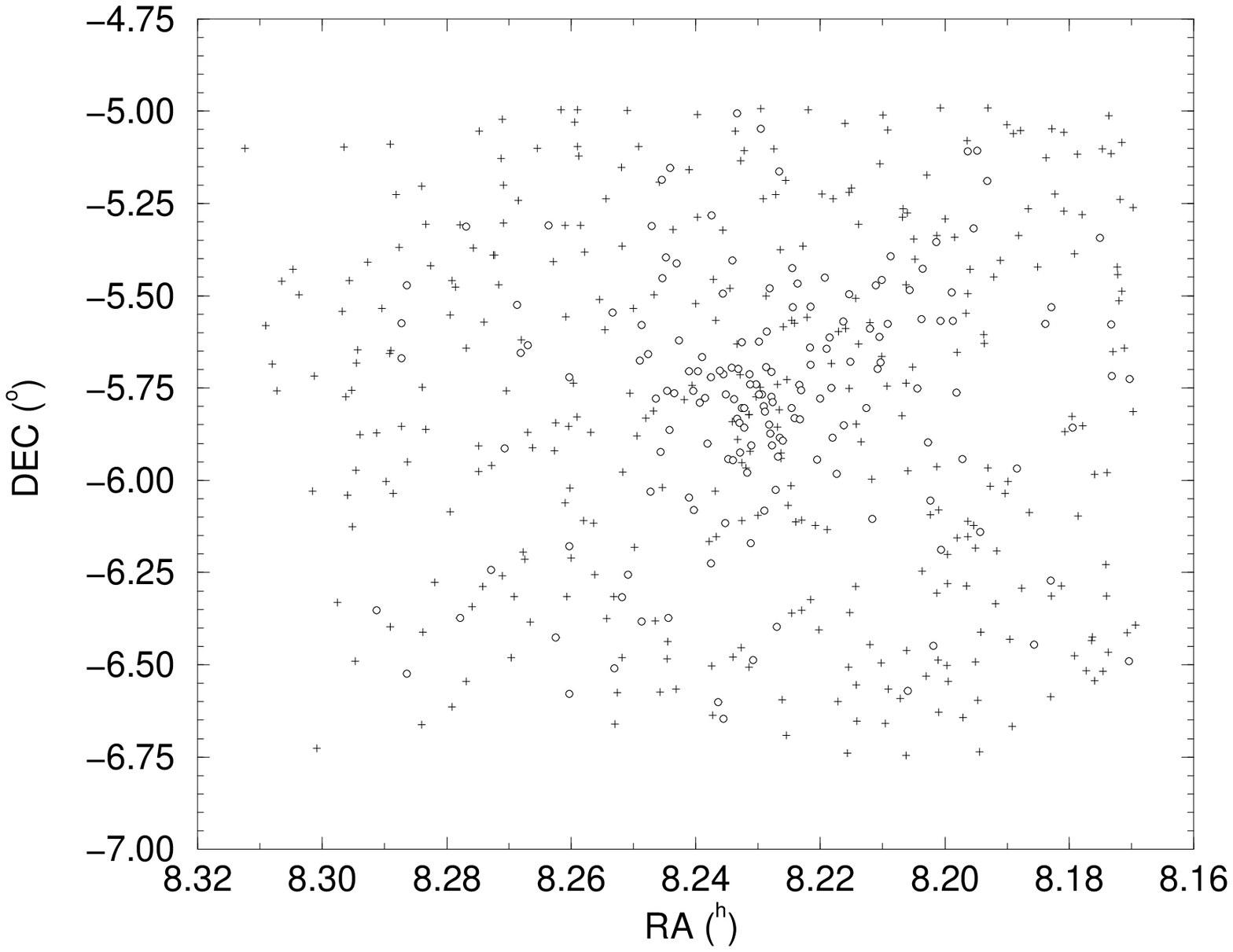}}
\caption{The position distribution of stars in NGC~2548 area.
         (``$\circ$" denotes a member of NGC~2548 with $P_c\ge0.7$,
          ``$+$" a field star with $P_c<0.7$)}
\label{coor}
\end{figure*}

   Figures~\ref{mvpr} and \ref{coor} show the proper motion vector-point
   diagram and the position distribution on the sky for all the measured
   stars respectively, where $``\circ"$ denotes a member of NGC~2548 with
   $P_c\ge0.7$, and all other stars are considered field stars indicated
   by ``$+$". 

\begin{figure}
\resizebox{6cm}{!}{\includegraphics{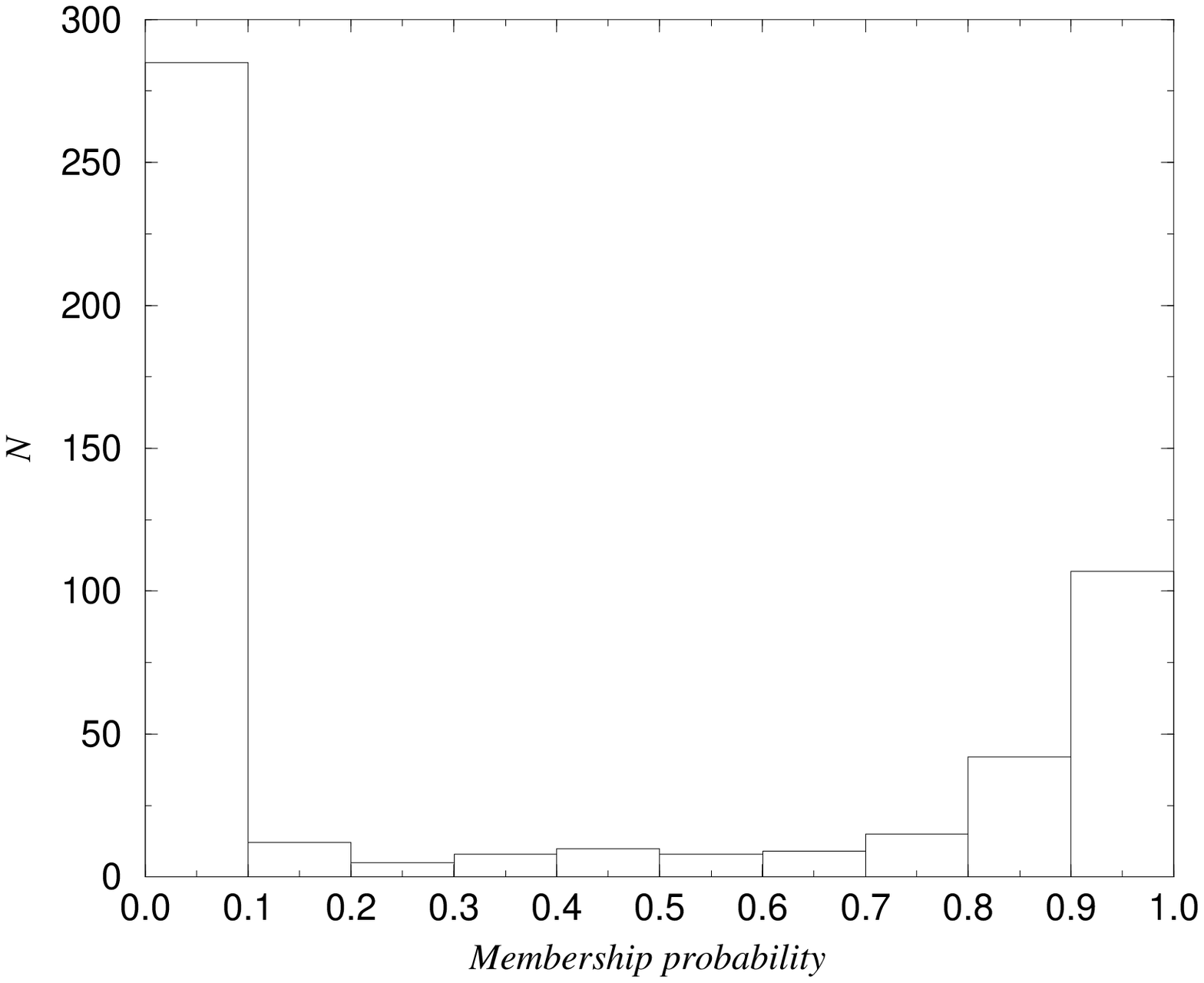}}
\caption{The histogram of cluster membership probability of NGC~2548}
\label{fig7}
\end{figure}

    The cluster membership probability histogram (Fig.~\ref{fig7}) shows a very
    clear separation between cluster members and field stars. The
    number of stars with membership probabilities higher than 0.7 for
    NGC~2548 is 165. The average cluster membership probability is 0.90, giving a
    contamination by field stars not larger than $10\%$. 
    If these stars were considered as belonging to the field, the apparent
    deficiency of field stars in the central region would vanish.
    Likewise, the contamination
    by cluster stars in the field is not larger than $6\%$.

    The present results for this cluster should be complemented with 
    photometry and radial velocity studies. To this end, our group 
    is working on a complete photometric study of the region.


\subsection{Effectiveness of membership determination}

    Contamination by background and foreground objects due to the
    observational projection effect cannot be avoided.
    Following Shao \& Zhao (\cite{sha}), we can assess the effectiveness of
    our membership determination, which
    is set as:

\begin{displaymath} E=1- \frac{N \sum_{i=1}^{N} \left\{ P_c(i)\left[1-P_c(i) \right]
\right\}} {\sum_{i=1}^{N} P_c(i) \sum_{i=1}^{N}
\left[1-P_c(i)\right] }\ ,  \end{displaymath}

\noindent where the greater $E$, the more effective the membership
determination.

    We thus obtained an effectiveness of membership
    determination of 0.77 for NGC~2548. Figure~3 of
    Shao $\&$ Zhao (\cite{sha}) paper shows that the effectiveness
    of membership determination of 43 open clusters ranges from
    0.20 to 0.90 and the peak value is 0.55. Compared with previous
    studies (Shao $\&$ Zhao \cite{sha}; Tian et al.\ \cite{tianzha};
    Balaguer-N\'u\~nez et al.\ \cite{Bai}), the effectiveness
    of membership determination for this open cluster is significantly high.

\subsection{Space velocity and galactic orbit}

    By combining our absolute proper motion results with
    radial velocity (Piatti et al.\ \cite{pia}), and the
    age and distance of the cluster from the Catalogue of Lyng\aa \
    (\cite{lynga}),
    we determined the space velocity and the galactic orbit of NGC~2548.

    To obtain the velocity of the cluster in the
    galactocentric frame, we assumed the motion of the Sun in the LSR
    to be $(U,V,W)_{\sun}$ = $(9.7,5.2,6.7)$ km~s$^{-1}$ (Bienaym\'e \cite{bien}).
    We adopted the current IAU standard values of $V_{LSR}$ = 220 km~s$^{-1}$
    for the local circular rotation velocity and $R_{\sun}$ = 8.5 kpc
    for the Sun galactocentric distance. The resulting
    galactocentric position and velocity of the cluster are
    $(x,y,z)$ = $(-8.89,-0.44,+0.16)$ kpc and $(U,V,W)$ = $(0,222,7)$ km~s$^{-1}$.
    These vectors, together with the galactic gravitational potential model
    of Allen \& Santillan (\cite{allen}) with the estimated cluster age of $\log t$~=~8.5,
    determine the orbit of the cluster in the Galaxy. The
    resulting orbit is characterized by an eccentricity of 0.022, radial oscillations
    between distances from the galactic center of 8.67 and 9.04
    kpc and vertical oscillations between $-0.19$ and 0.19 kpc. The
    cluster has made 1.18 revolutions around the galactic center
    and 8 crossings of the galactic disk in its lifetime. 
    Assuming the distance of Clari\'a (\cite{Claria}) does not affect the results.
    We currently observe the cluster near its maximum distance from the plane. This is 
    consistent with the expectation of a higher probability of finding the cluster near 
    its maximum $|z|$, where the cluster will spend more time due to its vertical motion.

\section{Summary}
    
    Proper motions and their corresponding errors of 501 stars within 
    a 1\fdg6 $\times$ 1\fdg6 area in
    the NGC~2548 region were determined from automatic MAMA measurements
    of 10 plates, five of which were newly taken. Comparison with Tycho-2 
    and Hipparcos Catalogues shows good agreement and underlines the precision
    of the proper motions derived in this paper. These proper motions are
    then used to determine membership probabilities of the stars in the region.
    We obtained 165 stars with membership pro\-ba\-bi\-li\-ties higher than 0.7.

\vspace{3mm}
%

\

\begin{acknowledgements}
    We would like to thank Prof. J.J.Wang for useful discussions
    and Dr.D.Galad\'{\i}-Enr\'{\i}quez for 
    help and useful comments. We would also like to thank
    R. Chesnel and P. Toupet for scanning and prereduction of the plates.
    This study was partially supported by the National Natural Science
    Foundation of China Grant No.19733001 and by 
    Joint Laboratory for Optional Astronomy
    of CAS. L.B-N. gratefully acknowledges financial support from Wang Kuan
    Cheng Fund, Chinese Academy of Sciences. This study was also partially
    supported by the contract No. A\&A2000-0937 with MCYT.

\end{acknowledgements}

\end{document}